\documentclass[aps,prc,showpacs,twocolumn,amssymb,floatfix]{revtex4}
\usepackage{graphicx}
\usepackage{color}

\newcommand{\be}{\begin{equation}}
\newcommand{\ee}{\end{equation}}
\newcommand{\bea}{\begin{eqnarray}}
\newcommand{\eea}{\end{eqnarray}}

\begin{document}
\title{Relativistic two-phonon model for low-energy nuclear response}
\author{Elena Litvinova}
\affiliation{Department of Physics, Western Michigan University,
Kalamazoo, MI 49008-5252, USA} \affiliation{National Superconducting
Cyclotron Laboratory, Michigan State University, East Lansing, MI
48824-1321, USA}
\author{Peter Ring}
\affiliation{Physik-Department der Technischen Universit\"at
M\"unchen, D-85748 Garching, Germany}
\author{Victor Tselyaev}
\affiliation{Nuclear Physics Department, St. Petersburg State
University, 198504 St. Petersburg, Russia}

\date{\today}

\begin{abstract}
A two-phonon version of the relativistic quasiparticle time blocking
approximation introduces as a new class of many-body models for
nuclear structure calculations based on the covariant energy density
functional. As a fully consistent extension of the relativistic
quasiparticle random phase approximation, the relativistic
two-phonon model implies fragmentation of nuclear states over
two-quasiparticle and two-phonon configurations coupled to each
other. In particular, we show how the lowest two-phonon $1^-$ state,
identified as a member of the $[2^+\otimes 3^-]$ quintuplet, emerges
from the coherent two-quasiparticle pygmy dipole mode in vibrational
nuclei. The inclusion of the two-phonon configurations into the
model space allows a quantitative description of the positions and
the reduced transition probabilities of the lowest 1$^-$ states in
tin isotopes $^{112,116,120,124}$Sn as well as the low-energy
fraction of the dipole strength below the giant dipole resonance
without any adjustment procedures. The model is applied to the
low-lying dipole strength in neutron-rich $^{68,70,72}$Ni isotopes.
Recent experimental data for $^{68}$Ni are reproduced fairly well.
\end{abstract}

\maketitle

\section{Introduction}

The theoretical description of nuclear low-lying dipole strength remains
among the most important problems in nuclear structure and nuclear
astrophysics. In contrast to high-frequency oscillations which
evolve very smoothly with particle numbers, the nuclear response below
the particle emission threshold is quite irregular. This phenomenon
reflects a complicated interplay of various structure mechanisms and
provides a sensitive test for microscopic theories which include
complex many-body correlations. It has been pointed out that
the low-energy enhancement of the dipole strength may enhance the
radiative neutron capture rates in the r-process of
nucleosynthesis~\cite{GKS.04} with a considerable influence on the
elemental abundance distributions \cite{S.09}. Since the r-process
path is believed to occur in environments of the extreme neutron
densities, the role of the exotic nuclear structures becomes
exceedingly important \cite{LG.10,PVKC.07}. Recent studies of the
neutron capture rates beyond the standard approaches \cite{LLL.09}
have revealed that the rates are sensitive to the fine structure of
the dipole strength distribution around the neutron emission
threshold of neutron-rich nuclei. This emphasizes the necessity of
a precise knowledge about the low-energy nuclear response and
stimulates the effort from both theoretical and experimental sides.

Measurements of the dipole strength by means of high resolution
nuclear resonance
fluorescence~\cite{Rye.02,ZVB.02,SBB.06,SRB.07,SRT.08,SFH.08}
resolve the fine structure of the spectra below the neutron
threshold. Unique spectroscopic information about the low-energy
dipole strength in neutron-rich medium-mass and heavy nuclei have
been obtained in recent experiments with Coulomb
dissociation~\cite{Adr.05,KAB.07} and virtual photon
scattering~\cite{WBC.09}. This offers exciting opportunities for
constraining the next-generation microscopic nuclear structure
models designed to provide the missing information about exotic
nuclei and to help in the analysis of experimental data.

Measurements of the low-lying dipole strength above and below the
neutron threshold are usually performed with different nuclear
reactions, which have reduced sensitivity in the area around the
threshold. Therefore, besides a few exceptions
\cite{Tamii.11,Pol.12,zr90.12}, a correct comparison of the
calculated low-lying strength with the data is still problematic.
Nevertheless, it is generally agreed that the low-energy dipole
response corresponds to oscillations of the neutron skin with respect
to the isospin saturated core \cite{SAZ.13}, therefore the
resonance-like structure on the low-energy shoulder of the giant
dipole resonance (GDR) is often called pygmy dipole resonance (PDR).
Properties of this mode of excitation are of special interest as
they are found to be correlated with the neutron skin thickness and
with the symmetry energy \cite{P.06,K.07,C.10}, although the degree
of such correlations is not well established yet \cite{RN.10}.

The collectivity of the PDR is another subject of discussions.
Self-consistent relativistic (quasiparticle) random phase
approximation (RQRPA) \cite{VPRL.01,PRN.03} results mostly in a relatively
collective pygmy mode, in contrast to the results of the
non-relativistic approaches~\cite{TSG.07,TL.08}, which give usually only
incoherent particle-hole transitions below the GDR. On the other
hand, in Ref.~\cite{TE.06} it is pointed out that the collectivity
of the pygmy mode is restored within non-relativistic QRPA
calculations with Skyrme forces when a fully self-consistent scheme
is employed. Hence, exact self-consistency between the mean
field and the nuclear effective interaction is an essential
ingredient for a successful description of the low-lying dipole
strength.

The degree of collectivity is conventionally quantified by the
number of the particle-hole transitions which contribute to a
particular excited state. However, this makes sense only if the
approach is, by construction, confined by the particle-hole
configurations. This is the case for the QRPA, but not for the true
wave functions which have more complicated structure. Since the
pygmy dipole mode has essentially surface nature, it strongly mixes
with other surface modes, especially with the low-lying ones.
Therefore, for an adequate description of the PDR QRPA is not
sufficient and correlations beyond QRPA should be included. For the
relativistic case, a self-consistent extension of the RQRPA has been
developed in Ref. \cite{LRT.08} where the coupling between
quasiparticles (q) and vibrational modes (phonons) is included by
means of introducing 2q$\otimes$phonon configurations into the
two-nucleon propagators. Applications of this approach called
relativistic quasiparticle time blocking approximation (RQTBA) have
demonstrated the importance of the quasiparticle-phonon coupling for
the description of the PDR. It has been found that the pygmy mode,
arising in RQRPA basically as a single state of isoscalar character
near the particle emission threshold, is strongly fragmented over
many states in a broad energy region when the coupling to phonons is
included. As a result, the major fraction of the strength is located
well below the original position of the RQRPA pygmy dipole mode.
Compared to existing data, the RQTBA describes the general behavior
of the dipole strength below the neutron threshold very
reasonably~\cite{LRV.07,LRT.08,LRTL.09,M.12} and is able to identify
the gross structure peculiarities of the PDR, in particular, its
isospin splitting \cite{E.10}. However, in order to account for the
fine structure of the spectrum, more correlations should be included
in the microscopic model. For instance, from the investigations of
Refs. \cite{Bry.99,Rye.02} within the Quasiparticle-Phonon Model
(QPM) \cite{Sol.92} one can conclude that configurations of the
two-phonon and  three-phonon types play a noticeable role and it is,
therefore, desirable to extend the RQTBA with these configurations.
In particular, the model has to reproduce the lowest dipole state in
vibrational nuclei, which is identified as a member of the
quintuplet $[2^+_1\otimes3^-_1]$. It was predicted in
Refs.~\cite{Lip.71,VK.71} and observed in spectra of spherical
medium-mass nuclei. While the $1^-$ member of this multiplet is
accessible by photon scattering experiments
~\cite{WRZ.96,Bry.99,Pys.06,OEN.07} and, therefore, relatively well
studied, data on the entire multiplets have been reported in only a
few cases \cite{RJB.94,BGB.95}. QPM as well as "Q-phonon" approach
formulated in both bosonic \cite{PBCOZ.94} and fermionic
\cite{JSV.04,JSV.06} configuration spaces describe nuclear excited
states of vibrational character by means of interacting bosons. Such
models explain well the physical mechanisms of multiphonon
excitations, however, they rely on the empirical input for the
phonon constituents.

In the present work, we show, in particular, how two-phonon states
can be described in a self-consistent approach based on the
covariant density functional theory. The two-phonon version of the
quasiparticle time blocking approximation (QTBA) proposed in
Ref.~\cite{Tse.07} is employed to introduce correlations between the
two quasiparticles within the 2q$\otimes$phonon configurations of
the conventional RQTBA. In this model, as in the RQTBA, the
quasiparticle-phonon coupling and the pairing correlations are
treated on an equal footing and spreading of the nuclear excited
states over two-phonon configurations appears in the excitation
spectra in addition to the so-called Landau damping of the
two-quasiparticle states. We compare the low-lying dipole strength
distribution obtained within the two-phonon RQTBA (RQTBA-2) to that
obtained within the conventional RQTBA and show how states of a
qualitatively new two-phonon nature emerge from the single
highly-correlated pygmy dipole modes in vibrational nuclei.

This article is an extended version of the letter in \cite{LRT.10};
here more details of the theoretical method are presented and a more
comprehensive analysis of the results is performed.
\section{Formalism}
\begin{figure*}[ptb]
\begin{center}
\includegraphics*[scale=0.8]{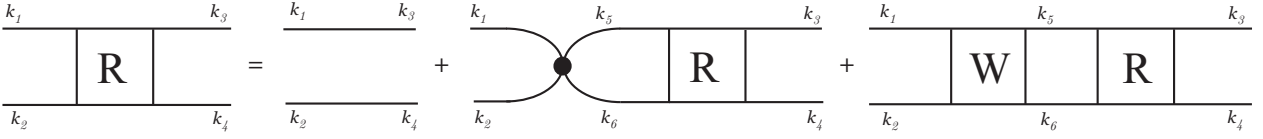}
\end{center}
\vspace{-0.5cm}
\caption{The Bethe-Salpeter equation for the response function $R$
of the many-body system in diagrammatic representation. The solid
lines with arrows denote single-quasiparticle propagators. The
integral part is divided into two terms, the small black circle
represents the static effective interaction $\tilde V$  and the
energy-dependent block $W(\omega)$ contains the dynamic
contributions induced by the coupling to phonons.}
\label{fig1}%
\vspace{-0.3 cm}
\end{figure*}
The response of finite Fermi-systems with even particle number to an
external perturbation is quantified by the response function which
describes the propagation of the two-quasiparticles in the
medium. The exact propagator includes, ideally, all possible kinds of the
in-medium interaction between two arbitrary quasiparticles and
contains all the information about the Fermi system that can be, in
principle, extracted by various experimental probes whose
interaction with the system can be represented by a
single-quasiparticle operator and can be taken into account in the
first order.

The response function $R$ is conventionally described by the
Bethe-Salpeter equation (BSE). The general form of this equation
\bea
R(14,23) = G(1,3)G(4,2) - \nonumber\\
- i\sum\limits_{5678}G(1,5)G(6,2)U(58,67)R(74,83),
\label{bse0}%
\eea
involves the one-nucleon Green's function (propagator) $G(1,2)$ in
the nuclear medium and the effective nucleon-nucleon interaction
$U(14,23)$. To include pairing correlations we use the formalism of
the extended (doubled) space of the single-particle states described
in Ref.~\cite{LRT.08}. So, the number indices $1,2,...$ include all
single-quasiparticle variables in an arbitrary representation in
this doubled space and time. Respectively, the summation over the
number indices implies an integration over the time variables. If $G$
is the exact single-quasiparticle Green's function, $U$ is the
amplitude of the effective interaction irreducible in the
$ph$-channel. This amplitude is determined as a variational
derivative of the nucleonic self-energy $\Sigma$ with respect to the
exact single-quasiparticle Green's function:
\begin{equation}
U(14,23)=i\frac{\delta\Sigma(4,3)}{\delta G(2,1)}. \label{uampl}%
\end{equation}
Similar to Refs. \cite{Tse.07,LRT.08}, we decompose both the
single-quasiparticle self-energy $\Sigma$ and the irreducible
effective interaction $U$ into static ${\tilde\Sigma}, \tilde V$ and
time-dependent $\Sigma^{(e)}, U^{(e)}$ parts as
\bea
\Sigma = {\tilde\Sigma} + \Sigma^{(e)}\\
U = {\tilde V} + U^{(e)}.
\eea Accordingly, we introduce the uncorrelated response ${\tilde
R}^{(0)}(14,23)={\tilde G}(1,3){\tilde G}(4,2)$, where ${\tilde
G}(1,2)$ are the single-quasiparticle mean-field Green functions in
the absence of the term $\Sigma^{(e)}$ in the self-energy.
Then, it can be shown that the BSE
(\ref{bse0}) takes the form:
\bea
R(14,23) = {\tilde G}(1,3){\tilde G}(4,2) - \nonumber\\
- i\sum\limits_{5678}{\tilde G}(1,5){\tilde G}(6,2)V(58,67)R(74,83),
\label{bse1}%
\eea
where $V$ is the new effective interaction amplitude which is
specified below. The well-known quasiparticle random phase
approximation (QRPA) including its relativistic version (RQRPA)
corresponds to the case of $V\approx{\tilde V}$ neglecting the
time-dependent term $U^{(e)}$. More precisely, in the (R)QRPA the
time-dependent term is included in a static approximation by
adjusting the parameters of the effective interaction $\tilde V$ to
ground state properties of nuclei such as masses and radii.
In the self-consistent (R)QRPA the static effective interaction is
nothing but the second variational derivative of the covariant
energy density functional (CEDF) $E[{\cal R}]$ with respect to the
density matrix $\cal R$ \cite{VALR.05}:
\be
{\tilde V}(14,23) = \frac{2\delta^2 E[{\cal R}]}{\delta{\cal
R}(2,1)\delta{\cal R}(3,4)} . \label{2-deriv}
\ee

In the approaches beyond the QRPA both static and time-dependent
terms are contained in the residual interactions. In medium-mass and
heavy nuclei vibrational and rotational modes emerge as collective
degrees of freedom which are strongly coupled to the single-particle
ones. The coupling to low-lying vibrations is known already for decades
\cite{BM.75} as a very important mechanism of the formation of nuclear
excited states and serves as a foundation for the so-called
(quasi)particle-phonon coupling model. Implementations of this
concept on the base of the modern density functionals have been
extensively elaborated in non-relativistic
\cite{CB.01,TSG.07,TSK.09,CSB.10,NCBBM.12} and relativistic
\cite{LRT.08,LRT.10,LA.11,L.12,MLVR.12} frameworks.

In the present work we consider the quasiparticle-phonon coupling
model for the time-dependent part of the nucleonic self-energy
$\Sigma^{(e)}$ and the effective interaction $U^{(e)}$. We consider the
BSE (\ref{bse1}) in the time blocking approximation, first proposed
in Ref. \cite{Tse.07} for superfluid Fermi systems and elaborated in
detail in Ref. \cite{LRT.08} for the relativistic framework. This
approximation allows an exact summation of a selected class of
Feynman's diagrams which give the leading contribution of the
quasiparticle-phonon coupling effects. To apply this method it is
convenient to write the BSE (\ref{bse1}) in the representation in
which the mean-field Green function $\tilde{G}$ is diagonal. In our
case (see Ref.~\cite{LRT.08} for more details) this representation
is given by the set of the eigenfunctions
$|\psi_{k}^{(\eta)}\rangle$ of the Relativistic Hartree-Bogoliubov
(RHB) Hamiltonian $\mathcal{H}_{RHB}$ satisfying the equations
\cite{KuR.91}:
\begin{equation}
\mathcal{H}_{RHB}|\psi_{k}^{(\eta)}\rangle=\eta E_{k}|\psi_{k}^{(\eta)}%
\rangle,\ \ \ \ {\cal H}_{RHB} = 2
\frac{\delta E[{\cal R}]}{\delta{\cal R}},
\label{hb}%
\end{equation}
where $E_{k}>0$, the index $k$ stands for the set of the
single-particle quantum numbers including states in the Dirac sea,
and the index $\eta = \pm 1$ labels positive- and negative-frequency
solutions of Eq.~(\ref{hb})
in the doubled quasiparticle space.
The eigenfunctions $|\psi_{k}^{(\eta)}\rangle$ are 8-dimensional
Bogoliubov-Dirac spinors of the following structure:
\begin{equation}
|\psi_{k}^{(+)}(\mbox{\boldmath $r$})\rangle=\left(
\begin{array}
[c]{c}%
U_{k}(\mbox{\boldmath $r$})\\
V_{k}(\mbox{\boldmath $r$})
\end{array}
\right)  ,\ \ \ \ |\psi_{k}^{(-)}(\mbox{\boldmath
$r$})\rangle=\left(
\begin{array}
[c]{c}%
V_{k}^{\ast}(\mbox{\boldmath $r$})\\
U_{k}^{\ast}(\mbox{\boldmath $r$})
\end{array}
\right)  . \label{dbasis}%
\end{equation}

Within the time blocking approximation mentioned above and after
performing a Fourier transformation to the energy domain, the BSE
(\ref{bse1}) for the spectral representation of the nuclear response
function $R(\omega)$ in the basis $\{|\psi_{k}^{(\eta)}\rangle\}$
reads:
\bea
R_{k_{1}k_{4},k_{2}k_{3}}^{\eta\eta^{\prime}}(\omega)=\tilde{R}_{k_{1}k_{2}%
}^{(0)\eta}(\omega)\delta_{k_{1}k_{3}}\delta_{k_{2}k_{4}}\delta^{\eta
\eta^{\prime}}+\nonumber\\
+ \tilde{R}_{k_{1}k_{2}}^{(0)\eta}(\omega)
\sum\limits_{k_{5}k_{6}\eta^{\prime\prime}}{V}_{k_{1}k_{6}%
,k_{2}k_{5}}^{\eta\eta^{\prime\prime}}(\omega)R_{k_{5}k_{4},k_{6}k_{3}}%
^{\eta^{\prime\prime}\eta^{\prime}}(\omega).
\label{respdir}%
\eea
The quantity ${\tilde R}^{(0)}$
\be
{\tilde R}^{(0)\eta}_{k_1k_2}(\omega) = \frac{1}{\eta\omega -
E_{k_1} - E_{k_2}}
\ee
describes the free propagation of two quasiparticles with their
Bogoliubov's energies $E_{k_1}$ and $E_{k_2}$ in the relativistic
mean field.
The interaction amplitude of Eq. (\ref{respdir}) contains both
static $\tilde V$ and dynamical (frequency-dependent) $W(\omega)$
parts (which will be specified below) and reads:
\bea
{V}_{k_{1}k_{4},k_{2}k_{3}}^{\eta\eta^{\prime}}(\omega)=\tilde{V}%
_{k_{1}k_{4},k_{2}k_{3}}^{\eta\eta^{\prime}} + {W}
_{k_{1}k_{4},k_{2}k_{3}}^{\eta\eta^{\prime}}(\omega), \nonumber \\
{W}_{k_{1}k_{4},k_{2}k_{3}}^{\eta\eta^{\prime}}(\omega) = \Bigl[\Phi_{k_{1}k_{4},k_{2}%
k_{3}}^{\eta}(\omega)-\Phi_{k_{1}k_{4},k_{2}k_{3}}^{\eta}(0)\Bigr]\delta%
^{\eta\eta^{\prime}}.
\label{W-omega}%
\eea

The diagrammatic representation of the Eq. (\ref{respdir}) is given
in Fig. \ref{fig1}. Notice here, that
in outward appearance this diagrammatic equation written, as in Ref.
\cite{LRT.08}, for the system with pairing correlations has the same
form as that for the normal (non-superfluid) system. The formal
similarity of the equations for the normal and superfluid systems is
achieved by the use of the representation of the basis functions
$|\psi_{k}^{(\eta)}\rangle$ satisfying Eq.~(\ref{hb}). This basis is
a counterpart of the particle-hole basis of the conventional RPA in
which the (Q)RPA equations have the most simple and compact form. In
the representation of the functions $|\psi_{k}^{(\eta)}\rangle$
the generalized superfluid mean-field Green function $\tilde{G}$
(often called Gor'kov-Green's function) has a diagonal form
and describes the propagation of the quasiparticle with the fixed energy.
In this diagonal representation the directions of the arrows in
the fermion lines of the diagrams (of the type shown in Fig. \ref{fig1})
label the positive- or the negative-frequency components of the functions
$\tilde{G}$. It should be noted that the so-called backward-going
diagrams corresponding to the ground-state correlations in the RQRPA
are not shown in Fig. \ref{fig1} though they are included in
Eq. (\ref{respdir}).

\begin{figure}[ptb]
\begin{center}
\includegraphics[scale=1.05]{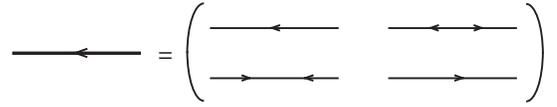}
\end{center}
\vspace{-0.5cm}
\caption{The 4-component Green's function in the diagrammatic
representation.}
\label{4-fg}%
\vspace{-0.3 cm}
\end{figure}

If we come back to the coordinate representation using
Eqs. (\ref{dbasis}) we get the non-diagonal Green function $\tilde{G}$
for the quasiparticle which has no definite energy.
In the diagrammatic language this Green function can be represented
by the 2$\times$2 block matrix shown in Fig. \ref{4-fg}.
The matrix elements of this matrix are the normal and anomalous
Green functions of the conventional Gor'kov theory.
In this representation it is clearly seen that
the difference between the normal
and the superfluid systems is that in the latter case all the quantities
acquire additional components in the doubled quasiparticle space.
Obviously, all these components are incorporated
in the representation of the functions $|\psi_{k}^{(\eta)}\rangle$
in the implicit form. On the other hand, the use of the basis
$\{|\psi_{k}^{(\eta)}\rangle\}$
allow us to reduce by a factor of 2 the dimension of the system of the
equations for the response function.
This property of the diagonal representation of the superfluid mean-field
Green functions has been already utilized in our previous papers
(see, e.g., Refs. \cite{LRT.08,Tse.07,LT.07}) but not discussed in
detail.

%
%

\begin{figure*}[ptb]
\begin{center}
\includegraphics*[scale=0.8]{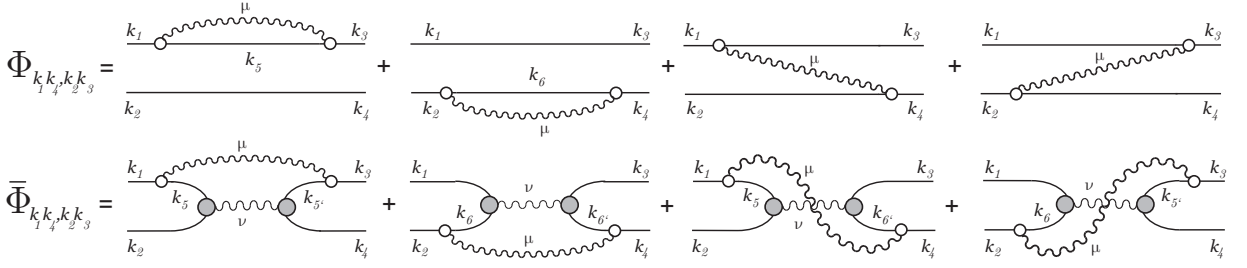}
\end{center}
\caption{The correspondence between the 2q$\otimes$phonon amplitude
$\Phi$ of the conventional phonon coupling model and the two-phonon
amplitude $\bar{\Phi}$ of the two-phonon model in a diagrammatic
representation. Solid lines with arrows and Latin indices denote the
single-quasiparticle nucleonic propagators, wavy curves with Greek
indices show the phonon propagators, empty circles represent phonon
vertices, and grey circles with the two attached nucleonic lines
denote the RQRPA transition densities (see text).}
\label{f1}%
\end{figure*}

In the present work, as in Refs.~\cite{LRT.08,LRTL.09}, the static
amplitude $\tilde V$ of Eq. (\ref{2-deriv}) is derived from the
covariant energy density functional $E[{\cal R}]$ with the parameter
set NL3~\cite{NL3} as a one-meson exchange interaction with a
non-linear self-coupling between the mesons. As in Ref. \cite{LRT.08}, pairing correlations
are introduced into the relativistic energy density functional as an
independently parameterized term in the form of a monopole interaction
with constant $G$ matrix elements. In the BSE the set of the quasiparticle
variables is doubled and therefore we have a formulation in
terms of 4-component Green's functions, as explained above. In
the 2q$\otimes$phonon version of the RQTBA the frequency-dependent
residual interaction $\Phi(\omega)$ is related to the
quasiparticle-phonon coupling self-energy by the consistency
condition and calculated within the quasiparticle time blocking
approximation. This approximation means that, due to the time
projection in the integral part of the BSE, the two-body propagation
through states with a more complicated structure than
2q$\otimes$phonon is blocked~\cite{Tse.07}. The diagrammatic
representation of the 2q$\otimes$phonon amplitude is shown in the
upper line of Fig. \ref{f1}.

In order to go a step further, notice, that in the time blocking
approximation the energy-dependent resonant part of the
two-quasiparticle amplitude $\Phi (\omega)$ can be factorized in a
special way \cite{Tse.07}. Namely, the two-quasiparticle
intermediate propagator, appearing as the two uncorrelated
quasiparticle lines between the phonon emission and absorption
vertices in the upper part of the Fig. \ref{f1}, can be extracted.
Thus, in the relativistic QTBA the amplitude $\Phi (\omega)$ takes
the following form:
\be
\Phi^{\eta}_{k_1k_4,k_2k_3} (\omega) = %
\sum_{k_5k_6,\mu} \zeta^{\,\mu\eta}_{k_1k_2;k_5k_6}\,
\tilde{R}^{(0)\eta}_{k_5k_6} (\omega - \eta\,\Omega_{\mu})\,
\zeta^{\,\mu\eta *}_{\,k_3k_4;k_5k_6}\,, \label{phires} \ee
where $\tilde{R}^{(0)\eta}_{k_5k_6}(\omega-\eta\,\Omega_{\mu})$ are
the matrix elements of the two-quasiparticle propagator in the mean
field with the frequency shifted forward or backward by the phonon
energy $\Omega_{\mu}$,
\bea
\zeta^{\,\mu(+)}_{\,k_1k_2;k_5k_6} =
\delta^{\vphantom{(+)}}_{k_1k_5}\,\gamma^{(-)}_{\mu;k_6k_2}
-\gamma^{(+)}_{\mu;k_1k_5}\delta^{\vphantom{(+)}}_{k_6k_2},
%
\nonumber \\
\zeta^{\,\mu(-)}_{\,k_1k_2;k_5k_6} =
\delta^{\vphantom{(+)}}_{k_5k_1}\,\gamma^{(+)\ast}_{\mu;k_2k_6}
-\gamma^{(-)\ast}_{\mu;k_5k_1}\delta^{\vphantom{(+)}}_{k_2k_6},
\label{zetapm}
\eea
so that
\be
\zeta^{\,\mu(-)}_{\,k_1k_2;k_5k_6}=%
-\zeta^{\,\mu(+)\ast}_{\,k_2k_1;k_6k_5}.
\ee
%
%

In Eq. (\ref{zetapm}) and below we use a shorthand notation for
the phonon amplitudes involved in the present
work:
\begin{equation}
\gamma^{\eta}_{\mu;k_1k_2} =
\gamma^{\eta_1\eta_2}_{\mu;k_1k_2}\delta_{\eta\eta_1}\delta_{\eta\eta_2},
\ \ \ \ \eta = (\pm)
\end{equation}
where the vertices ${\gamma}_{\mu;k_{1}k_{2}}^{\eta_{1}\eta_{2}}$
determine the coupling of the quasiparticles to the collective
vibrational state (phonon) $\mu$.
%
In the conventional version of the
quasiparticle-vibration coupling model these vertices are derived
from the corresponding RQRPA transition densities
$\mathcal{R}_{\mu}$ and the static effective interaction as
\begin{equation}
\gamma_{\mu;k_{1}k_{2}}^{\eta_{1}\eta_{2}}=\sum\limits_{k_{3}k_{4}}%
\sum\limits_{\eta}
\tilde{V}_{k_{1}k_{4},k_{2}k_{3}}^{\eta_{1},%
-\eta_,\eta_{2},\eta}\mathcal{R}_{\mu;k_{3}k_{4}}^{\eta}
\label{phonon}%
\end{equation}
where
$\tilde{V}_{k_{1}k_{4},k_{2}k_{3}}^{\eta_1\eta_4,\eta_2\eta_3}$ is
the matrix element of the amplitude $\tilde{V}$ of
Eq.~(\ref{2-deriv}) in the basis $\{|\psi_{k}^{(\eta)}\rangle\}$.
The matrix elements of the phonon transition densities are
calculated, in a first approximation, within the relativistic
quasiparticle random phase approximation \cite{PRN.03}. In the
Dirac-Hartree-BCS basis $\{|\psi_{k}^{(\eta)}\rangle\}$ it has the
following form:
\begin{equation}
\mathcal{R}_{\mu;k_{1}k_{2}}^{\eta}={\tilde{R}}_{k_{1}k_{2}}^{(0)\eta}%
(\Omega_{\mu})\sum\limits_{k_{3}k_{4}}\sum\limits_{\eta^{\prime}}\tilde
{V}_{k_{1}k_{4},k_{2}k_{3}}^{\eta\eta^{\prime}}\mathcal{R}_{\mu;k_{3}k_{4}%
}^{\eta^{\prime}}, \label{rqrpa}%
\end{equation}
where
$\tilde{V}_{k_{1}k_{4},k_{2}k_{3}}^{\eta\eta^{\prime}}=\tilde{V}_{k_{1}%
k_{4},k_{2}k_{3}}^{\eta,-\eta^{\prime},-\eta,\eta^{\prime}}$.
This means that we cut out the components of the tensors in the
quasiparticle space, which are relevant for the particle-hole
channel.

\begin{figure}
\begin{center}
\includegraphics*[scale=0.4]{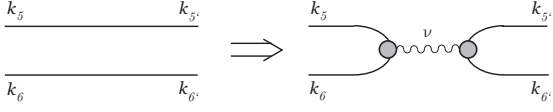}
\end{center}
\caption{Replacement of the uncorrelated two-nucleon propagator by
the correlated one. Grey circles with the two attached nucleonic
lines and wavy curve denote the RQRPA transition densities and the
phonon propagator, respectively.}
\label{subst}%
\end{figure}

In the graphic expression of the amplitude (\ref{phires}) in the
upper line of the Fig.~\ref{f1} the uncorrelated propagator
$\tilde{R}^{(0)\eta}_{k_1k_2}$ is represented by the two straight
nucleonic lines between the circles denoting emission and absorption
of the phonon by a single quasiparticle with amplitudes
$\gamma^{\eta_1\eta_2}_{\mu;k_1k_2}$. The approach to the amplitude
$\Phi(\omega)$ expressed by the Eq. (\ref{phires}) represents a
version of first-order perturbation theory compared to RQRPA and
the amplitude $\Phi(\omega)$ of Eq. (\ref{W-omega}) is the
first-order correction to the effective interaction ${\tilde V}$,
because the dimensionless matrix elements of the phonon vertices are
such that $\gamma^{\eta_1\eta_2}_{\mu;k_1k_2}/\Omega_{\mu}\ll 1$ in
most of the physical cases. The phonon-coupling term $\Phi$
generates fragmentation of nuclear excited states. For giant
resonances this fragmentation is the source of the spreading width
and in the low-energy region below the neutron threshold this term
is responsible almost solely for the appearing strength. In the
relativistic framework, this was confirmed and extensively studied
\cite{LRT.08,LRTL.09} and verified by comparison to experimental
data \cite{E.10,M.12}. However, a comparison with high-resolution
experiments on the dipole strength below the neutron threshold has
revealed that although the total strength and some gross features of
the strength are reproduced well, the fine features are sensitive to
the truncation of the configuration space by 2q$\otimes$phonon
configurations and further extensions of the method are due. The
first possible extension of this model uses the idea proposed in
Ref.~\cite{Tse.07}. It is based on the factorization of Eq.
(\ref{phires}): the uncorrelated propagator ${\tilde R}^{(0)\eta}$
in Eq. (\ref{phires}) is replaced by the positive- ($\eta=+1$) or
the negative- ($\eta=-1$) frequency part of a correlated one which,
in first order approximation, is the antisymmetrized RQRPA
propagator $R^{(RQRPA)\eta}$. Instead of the amplitude $\Phi$, we
have the new amplitude $\bar\Phi$:
\bea
&&{\bar\Phi}^{\eta}_{k_1k_4,k_2k_3} (\omega)
= \frac{1}{2}
\sum_{k_5k_6,k_{5^{\prime}}k_{6^{\prime}}\mu}
\zeta^{\,\mu\eta}_{k_1k_2;k_5k_6}\,
\nonumber\\%
&\times&
R^{(RQRPA)\eta}_{k_5k_{6^{\prime}},k_6k_{5^{\prime}}} (\omega -
\eta\,\Omega_{\mu})\, \zeta^{\,\mu\eta
*}_{\,k_3k_4;k_{5^{\prime}}k_{6^{\prime}}}.
\label{phires2}
\eea
The factor $1/2$ in Eq.~(\ref{phires2}) appears due to the
antisymmetrization of Eq. (\ref{phires}) as was explained
in Ref.~\cite{Tse.07}. This antisymmetrization implies that
the following equations are fulfilled
\bea
R^{(RQRPA)\eta}_{k_1k_4,k_2k_3} (\omega) &=& -
R^{(RQRPA)\eta}_{k_2k_4,k_1k_3} (\omega)
\nonumber\\
&=& -
R^{(RQRPA)\eta}_{k_1k_3,k_2k_4} (\omega)
\label{symrf}
\eea
which are not valid for the uncorrelated propagator ${\tilde
R}^{(0)\eta}$. By this substitution, we introduce RQRPA correlations
into the intermediate two-quasiparticle propagators, i.e., in
diagrammatic language, we perform the operation shown in Fig.
\ref{subst}.
Thus, two-phonon configurations appear in the amplitude $\Phi
(\omega)$ as it is clear from the diagrammatic representation of the
new amplitude $\bar\Phi$ shown in the bottom line of
Fig.~\ref{f1}. The analytic expression of this amplitude reads:
\be {\bar \Phi}^{\eta}_{k_1k_4,k_2k_3} (\omega) = \frac{1}{2}
\sum_{\mu,\nu}\frac{{\bar\zeta}^{\,\eta}_{\,\mu\nu;k_1k_2}\,
{\bar\zeta}^{\,\eta *}_{\,\mu\nu;k_3k_4}}
{\eta\,\omega-\Omega_{\mu}-\Omega_{\nu}}\,, \label{phiresc} \ee
where
\bea {\bar\zeta}^{\,(+)}_{\,\mu\nu;k_1k_2}&=&\sum_{k_6} {\cal
R}^{(+)}_{\,\nu;k_1k_6}\,\gamma^{(-)}_{\,\mu;k_6k_2}\, - \sum_{k_5}
\gamma^{(+)}_{\,\mu;k_1k_5}\,{\cal
R}^{(+)}_{\,\nu;k_5k_2}\nonumber\\
{\bar\zeta}^{\,(-)}_{\,\mu\nu;k_1k_2}&=&
{\bar\zeta}^{\,(+)\ast}_{\,\mu\nu;k_2k_1}, \eea
%
%
%
%
%
and ${\cal R}^{\eta}_{\,\nu;k_1k_2}$ are the matrix elements of the
RQRPA transition densities defined in Eq. (\ref{rqrpa}) and
corresponding to grey circles together with two nucleonic lines in
Fig.~\ref{f1}. One can show that in the limit of vanishing
static interaction $\tilde{V}$ between the two intermediate
quasiparticles Eq. (\ref{phiresc}) transforms to the antisymmetrized
Eq. (\ref{phires}) of the original (R)QTBA.
As in conventional (R)QTBA, the elimination of double counting
effects in the phonon coupling is performed by the subtraction of
the static contribution of the amplitude ${\bar \Phi}$ from the
residual interaction in Eq. (\ref{W-omega}), since the parameters of
the CEDF have been adjusted to experimental data for ground states
and include therefore already essential phonon contributions to the
ground state. Therefore, the BSE in the two-phonon model has the
same form as Eq. (\ref{respdir}), but it contains the amplitude
$\bar\Phi$ instead of $\Phi$.

The calculations are performed in the following four steps: (i) The
RHB equation (\ref{hb}) is solved to determine the
single-quasiparticle energies and wave functions. These wave
functions serve as a working basis for the subsequent calculations.
(ii) The phonon frequencies, their coupling vertices $\gamma^{\eta}$
and the transition densities ${\cal R}^{\eta}$ are calculated within
the self-consistent RQRPA using the static residual interaction
$\tilde V$. (iii) The BSE for the correlated propagator
$R^{(e)}(\omega)$
\bea R^{(e)\eta}_{k_1k_4,k_2k_3}(\omega) = {\tilde
R}^{(0)\eta}_{k_1k_2}(\omega)\delta_{k_1k_3}\delta_{k_2k_4} +\nonumber\\
+{\tilde R}^{(0)\eta}_{k_1k_2}(\omega)
\sum\limits_{k_5k_6}\bigl[{\bar\Phi}^{\eta}_{k_1k_6,k_2k_5}(\omega)
- {\bar\Phi}^{\eta}_{k_1k_6,k_2k_5}(0)\bigr]\times\nonumber\\
\times R^{(e)\eta}_{k_5k_4,k_6k_3}(\omega), \label{respdir2} \eea
is solved in the Dirac-Hartree-BCS basis; (iv) the BSE for the
full response function $R(\omega)$ \bea
R_{k_{1}k_{4},k_{2}k_{3}}^{\eta\eta^{\prime}}(\omega) &=&
R^{(e)\eta}_{k_1k_4,k_2k_3}(\omega)\delta^{\eta\eta^{\prime}}
\nonumber\\
&+& \sum\limits_{k_{5}k_{6}k_{7}k_{8}}
R^{(e)\eta}_{k_1k_6,k_2k_5}(\omega)
\nonumber\\
&\times&
\sum\limits_{\eta^{\prime\prime}}{\tilde V}_{k_{5}k_{8}%
,k_{6}k_{7}}^{\eta\eta^{\prime\prime}}R_{k_{7}k_{4},k_{8}k_{3}}%
^{\eta^{\prime\prime}\eta^{\prime}}(\omega)
\label{respdir1}%
%
\eea
where
\bea R_{k_{1}k_{4},k_{2}k_{3}}^{\eta\eta^{\prime}}(\omega)&=&
R_{k_{1}k_{4},k_{2}k_{3}}^{\eta,-\eta^{\prime},-\eta,\eta^{\prime}}(\omega),\nonumber\\
{\tilde V}_{k_{1}k_{4},k_{2}k_{3}}^{\eta\eta^{\prime}}&=&{\tilde
V}_{k_{1}k_{4},k_{2}k_{3}}^{\eta,-\eta^{\prime},-\eta,\eta^{\prime}},
\eea
%
%
is solved in the momentum-channel representation which is especially
convenient because of the structure of the one-boson exchange
interaction. The details are given below.

\section{Bethe-Salpeter equation in the coupled form}

For the spherically symmetric case Eq. (\ref{respdir2}) is
formulated in terms of the reduced matrix elements with the
transferred angular momentum $J$, i.e. in the so-called coupled
form, as follows:
\bea
R_{(k_{1}k_{4},k_{2}k_{3})}^{(e)J,\eta}(\omega)&=&{\tilde{R}}^{(s)J,\eta}%
_{(k_{1}k_4,k_{2}k_3)}(\omega)+\nonumber\\
+{\tilde{R}}^{(0)\eta}_{(k_{1}k_{2})}(\omega)
\sum\limits_{(k_{6}\leq k_{5})}\Bigl[{\bar\Phi}_{(k_{1}k_{6},k_{2}%
k_{5})}^{(s)J,\eta}(\omega)&-&{\bar\Phi}_{(k_{1}k_{6},k_{2}k_{5})}^{(s)J,\eta}%
(0)\Bigr]\times\nonumber\\
&\times&R_{(k_{5}k_{4},k_{6}k_{3})}^{(e)J,\eta}(\omega).
\label{correlated-propagator}%
\eea
The index "(s)" implies here that the corresponding matrix elements
are symmetrized with respect to one non-conjugated and one
conjugated quasiparticle index. Such a symmetrization allows a
shortened summation in the integral part of the Eq.
(\ref{correlated-propagator}) and simplifies, to some extent, the
numerical calculations. The symmetrized matrix elements of the mean
field propagator ${\tilde R}^{(s)}$ and of the two
quasiparticles-phonon coupling amplitude $\Phi^{(s)}$ have the
following form:
\begin{eqnarray}
{\tilde{R}}^{(s)J,\eta}_{(k_{1}k_{4},k_{2}k_{3})}(\omega)=%
{\tilde R}^{(0)\eta}_{(k_1k_2)}(\omega) \times\nonumber\\ \times
\bigl[\delta_{(k_1k_3)}\delta_{(k_2k_4)}+(-)^{\phi_{12}}%
\delta_{(k_1k_4)}\delta_{(k_2k_3)} \bigr],
\\
{\bar\Phi}^{(s)J,\eta}_{(k_1k_4,k_2k_3)}(\omega)=\frac{1}{1+%
\delta_{(k_3k_4)}} \times\nonumber\\ \times
\Bigl[{\bar\Phi}^{J,\eta}_{(k_1k_4,k_2k_3)}(\omega) +
(-)^{\phi_{12}}{\bar\Phi}^{J,\eta}_{(k_2k_4,k_1k_3)}(\omega)\Bigr],
\end{eqnarray}
where $\phi_{12}=J+l_1-l_2+j_1-j_2$. The reduced matrix elements of
the phonon coupling amplitudes for the two-phonon model can be
expressed as:
\be {\bar\Phi}^{J,\eta}_{(k_1k_4,k_2k_3)}(\omega) =
\frac{1}{2(2J+1)}\sum\limits_{(\mu,\nu)}
\frac{{\bar\zeta}^{\eta}_{(\mu\nu;k_1k_2)}{\bar\zeta}^{\eta\ast}_{(\mu\nu;k_3k_4)}}{\eta\omega
- \omega_{\mu} - \omega_{\nu}}\,, \label{phibar} \ee
\bea {\bar\zeta}^{(+)}_{(\mu\nu;k_1k_2)}=(-1)^{j_1+j_2} \times\nonumber\\
\times\Bigl[ \sum\limits_{(k_6)} \gamma_{({\mu};k_6k_2)}^{(-)}{\cal
R}_{({\nu};k_1k_6)}^{(+)} \left\{
\begin{array}
[c]{ccc}%
J_{\mu} & J_{\nu} & J\\
j_{1} & j_{2} & j_6%
\end{array}
\right\} \nonumber\\
-(-1)^{J_{\mu}+J_{\nu}+J}\sum\limits_{(k_5)}\gamma_{(\mu;k_1k_5)}^{(+)}{\cal
R}_{(\nu;k_5k_2)}^{(+)} \left\{
\begin{array}
[c]{ccc}%
J_{\nu} & J_{\mu} & J\\
j_{1} & j_{2} & j_5%
\end{array}
\right\} \Bigr],\nonumber
%
\eea
\be
{\bar\zeta}^{(-)}_{(\mu\nu;k_1k_2)}=(-1)^{J_{\mu}+J_{\nu}+J}{\bar\zeta}^{(+)}_{(\mu\nu;k_2k_1)}.
\ee
Then, the BSE for the full response function $R(\omega)$
\bea R_{(k_{1}k_{4},k_{2}k_{3})}^{J,\eta\eta^{\prime}}(\omega)=
R_{(k_{1}k_{4},k_{2}k_{3})}^{(e)J,\eta}(\omega)\delta^{\eta\eta^{\prime}}
+\sum\limits_{(k_{6}\leq k_{5})}\times\nonumber\\
\times\sum\limits_{(k_{8}\leq
k_{7})\eta^{\prime\prime}}R_{(k_{1}k_{6},k_{2}k_{5})}^{(e)J,\eta}(\omega)
{\tilde V}^{J,\eta\eta^{\prime\prime}}_{(k_{5}k_{8},k_{7}k_{6})}R_{(k_{7}%
k_{4},k_{8}k_{3})}^{J,\eta^{\prime\prime}\eta^{\prime}}(\omega)\nonumber\\
\label{responsej}%
\eea
is solved in both Dirac-Hartree-BCS and momentum-channel
representations. This latter representation is especially convenient
for numerical solution of the BSE with the static effective
interaction $\tilde V$ of the one-boson exchange type. The
expressions for the reduced matrix elements ${\tilde
V}^{J,\eta\eta^{\prime\prime}}_{(k_{5}k_{8},k_{7}k_{6})}$ of this
interaction in doubled quasiparticle space and further details
are given in the Appendix C of Ref. \cite{LRT.08}. The resulting
linear response function $R(\omega)$ contains all the information
on the nuclear response to external one-body operators. To describe the
observed spectrum of a nucleus excited by a sufficiently weak
external field $P$ as, for instance, an electromagnetic field, one
has to make a double convolution of the response function with this
field. The reduced matrix elements of the external field operator
have the following general coupled form:
\bea
P_{(k_{1}k_{2})}^{(p)J,\eta}=\sum\limits_{LS}
\frac{\delta_{\eta,1}+(-1)^{S}\delta_{\eta,-1}}{\sqrt{1 +
\delta_{(k_1k_2)}}}
\eta^{S}_{(k_1k_2)}\times\nonumber\\\times\langle(k_{1})\parallel
P^{(p)J}_{LS}\parallel(k_{2})\rangle,
\eea
where the index $(p)$ contains all possible quantum numbers other than
those concretized here. The factors $\eta^{S}_{(k_1k_2)}$ are the
conventional factors \cite{RS.80} determined by
combinations of the quasiparticle occupation numbers $u_k, v_k$:
\be
\eta_{(k_{1}k_{2})}^{S} =\frac{1}{\sqrt{1+\delta_{(k_{1}k_{2})}}%
}\Bigl(u_{k_{1}}v_{k_{2}}+(-1)^{S}v_{k_{1}}u_{k_{2}}\Bigr)\label{eta}
\ee
obtained as a solution of Eq. (\ref{hb}). Such combinations arise
due to symmetrization in the integral part of the Eq.
(\ref{responsej}), which enables one to take each $2q$-pair into
account only once because of the symmetry properties of the reduced
matrix elements ${\tilde
V}^{J,\eta\eta^{\prime\prime}}_{(k_{5}k_{8},k_{7}k_{6})}$.

The double convolution of the response function with the external
field operator defines the quantity called nuclear polarizability:
\bea
\Pi_{P}(\omega) &=&
\sum\limits_{(k_{2}\leq
k_{1})\eta}\sum\limits_{(k_{4}\leq
k_{3})\eta'}P_{(k_{1}k_{2})}^{(p)J,\eta\ast}
\nonumber\\
&\times&
R_{(k_{1}k_{4}%
,k_{2}k_{3})}^{J,\eta\eta'}(\omega)P_{(k_{3}k_{4})}^{(p)J,\eta'}
%
%
\eea
which determines the microscopic strength function $S(E)$ as:
\be S(E) = -\frac{1}{\pi}\lim\limits_{\Delta\to +0}Im
\Pi_P(E+i\Delta). \ee
In the calculations a finite imaginary part $\Delta$ of the energy
variable is introduced for convenience in order to obtain a smoothed
envelope of the spectrum. This parameter introduces an additional
artificial width for each excitation. This width emulates
effectively contributions from configurations higher than 4q and the
coupling to the continuum, which are not taken into account explicitly
in our approach.

\section{Results and discussion}

\begin{figure*}[ptb]
\begin{center}
\vspace{-3cm} \hspace{-0.93cm}
\includegraphics*[scale=0.75]{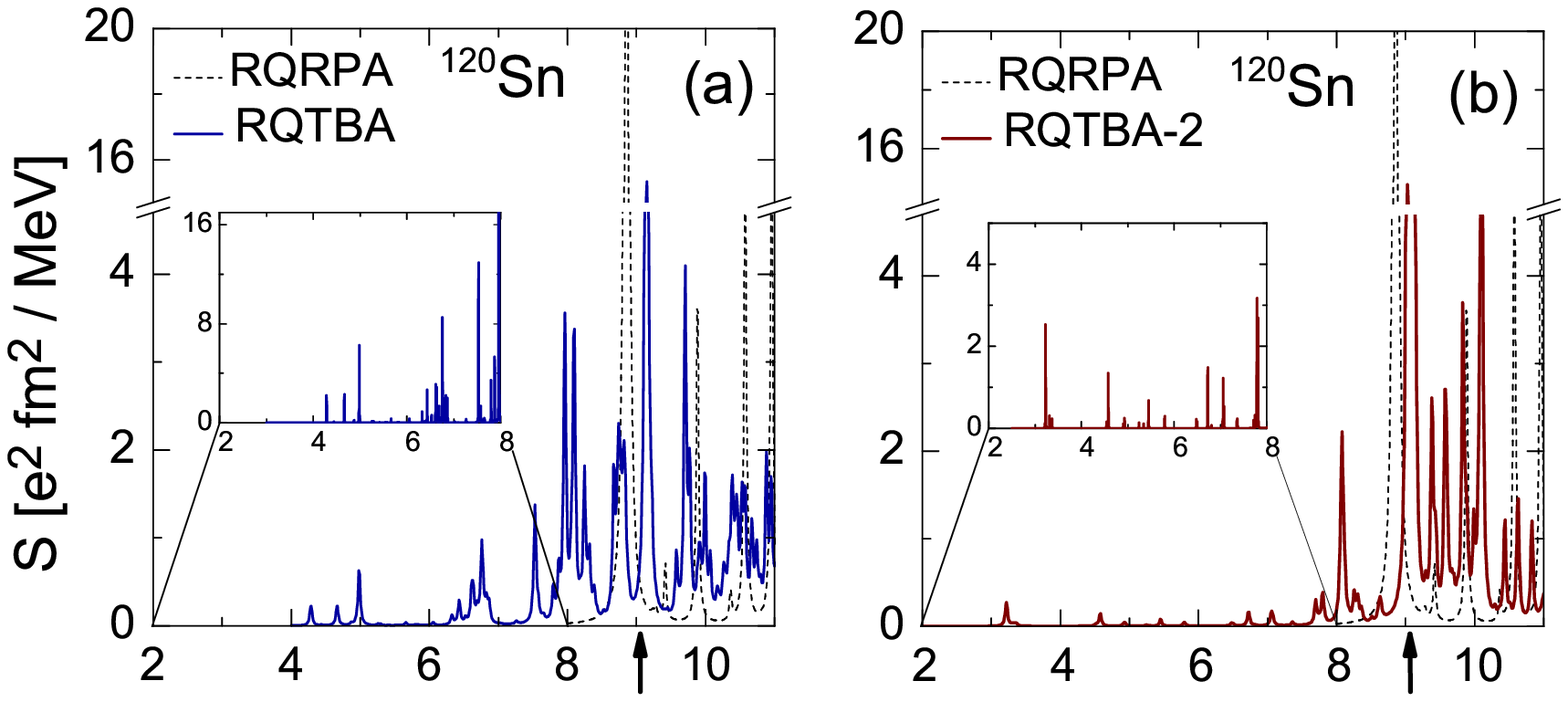}
\vspace{-1.9cm}
\end{center}
\begin{center}
\vspace{-3.0cm} \hspace{-0.93cm}
\includegraphics*[scale=0.75]{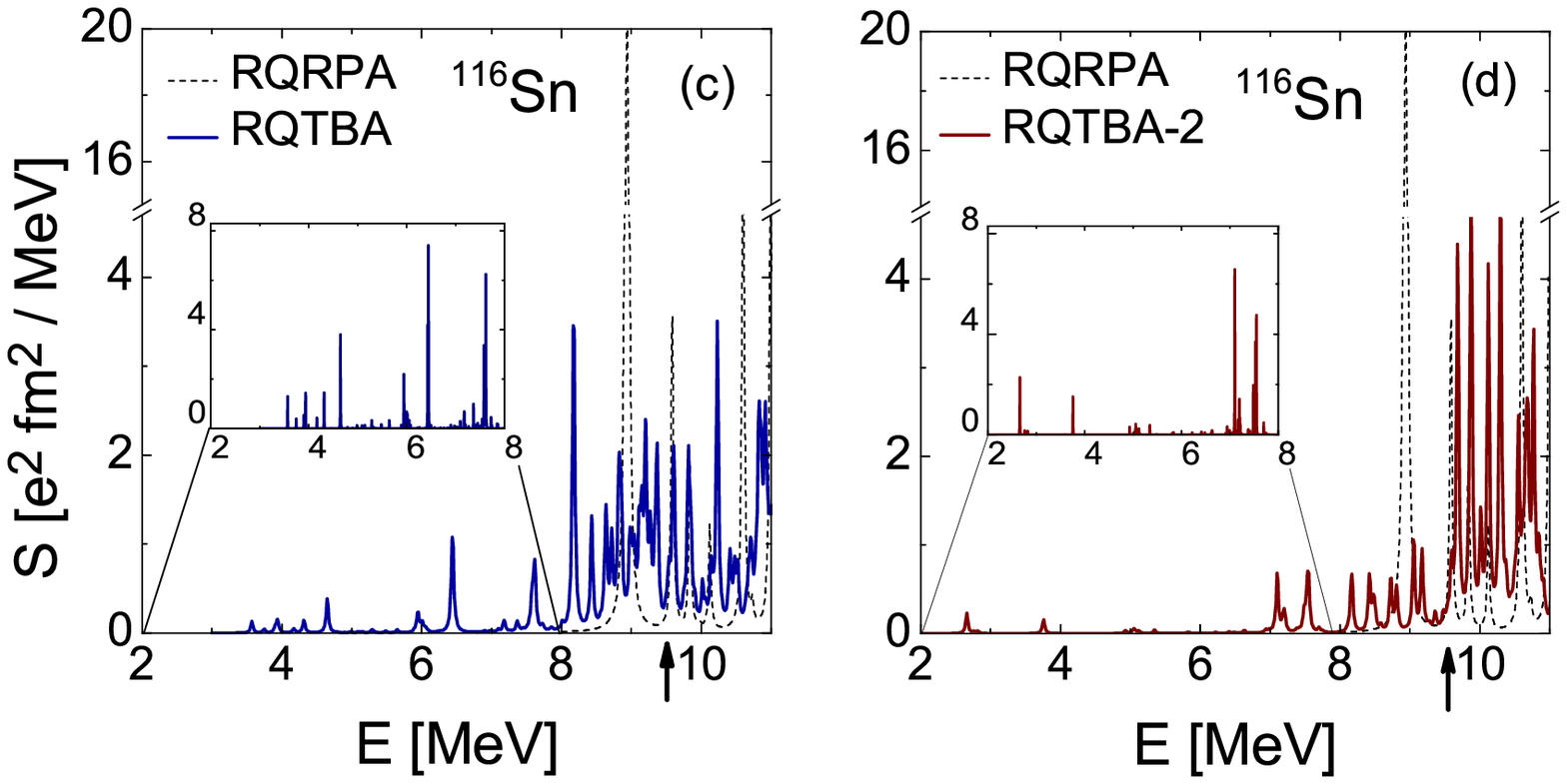}
\vspace{-0.8cm}
\end{center}
\caption{Low-lying dipole spectra of $^{116,120}$Sn calculated
within RQRPA (dashed curves), RQTBA (blue solid curve, panels
(a,c)) and RQTBA-2 (red solid curve, panels (b,d)). A finite
smearing parameter $\Delta$ = 20 keV has been used in the
calculations. The inserts show the zoomed pictures of the spectra
below 8 MeV with a smaller value $\Delta$ = 2 keV allowing to see
all the states in this energy region. The arrows indicate the
neutron thresholds.}
\label{fig3}%
\vspace{-0.2 cm}
\end{figure*}

To illustrate the effect of the two-phonon correlations on spectra
of nuclear excitations, we consider the dipole response of tin and
nickel isotopes in the energy region below the giant dipole
resonance (GDR). Fig.~\ref{fig3} displays the electromagnetic dipole
strength functions in $^{116,120}$Sn calculated within conventional
RQTBA~\cite{LRT.08} and two-phonon RQTBA-2 presented here. The
strength functions obtained in this way are compared with the
original RQRPA strength function because both of them originate from
RQRPA by different fragmentation mechanisms. The first observation
is that the total strength $\sum B(E1)\uparrow$ below the neutron
threshold is reduced in RQTBA-2. For example, for $^{116}$Sn we have
0.20 e$^2$ fm$^2$ below 8 MeV which agrees with 0.204(25) e$^2$
fm$^2$ obtained in the experiment of Ref.~\cite{Gov.98}. For
$^{120}$Sn, if we include the relatively strong state at 8.08 MeV
into the integration region, this quantity is 0.31 e$^2$ fm$^2$, in
agreement with the results of the quasiparticle phonon model (QPM)
of 0.289 e$^2$ fm$^2$~\cite{TLS.04}. One also notices that in both
nuclei the major fraction of the RQRPA pygmy mode shown by the
dashed curve is pushed up above the neutron threshold by the RQTBA-2
correlations.


\begin{table}[ptb]
\caption{The energies, reduced transition probabilities and
anharmonicities of the lowest 1$^-$ states in $^{112,116,120,124}$Sn
isotopes calculated in the relativistic two-phonon model, compared
to data of Refs.~\cite{Pys.06,Bry.99,OEN.07,Gov.98,Oezel.phd} and to
QPM calculations of Ref. \cite{Pys.06}, see text for details.}
\label{tab1}
\begin{center}
\tabcolsep=0.8em \renewcommand{\arraystretch}{1.0}%
\begin{tabular}
[c]{ccccc}\hline\hline
 &  & $\omega$(1$^-_1$) & B(E1)$\uparrow$ & $R_{\omega}$ \\
 &  & (MeV) & ($10^{-3}$ e$^2$ fm$^2$) &  \\
 \hline
 & RQTBA-2 & 3.85 & 26.12 &  0.98\\
 $^{112}$Sn & QPM  & 3.24 & 1.6 & \\
  & Exp. \cite{Pys.06} & 3.43 & 10.7(12) & 0.95 \\
  & Exp. \cite{OEN.07} & 3.43 & 15.0(10) & \\
 \hline
 & RQTBA-2 & 2.66 & 14.33& 0.94 \\
 $^{116}$Sn & QPM & 3.35 & 8.2 & \\
  & Exp. \cite{Bry.99} & 3.33 & 6.55(65) & 0.94 \\
  & Exp. \cite{Gov.98,OEN.07} & 3.33 & 16.3(9) &  \\
 \hline
 & RQTBA-2 & 3.23 & 15.90 & 0.95 \\
 $^{120}$Sn & QPM & 3.32 & 7.2 & \\
  & Exp. \cite{Bry.99} & 3.28 & 7.60(51) & 0.92 \\
  & Exp. \cite{Oezel.phd} & 3.28 & 11.2(11) &  \\
 \hline
 & RQTBA-2 & 3.98 & 12.91 &  0.97\\
 $^{124}$Sn & QPM & 3.57 & 3.5 & \\
  & Exp. \cite{Bry.99} & 3.49 & 6.1(7) & 0.93 \\
 & Exp. \cite{Gov.98,OEN.07} & 3.49 & 10.0(5) &  \\
 \hline\hline
\end{tabular}
\end{center}
\vspace{-7mm}
\end{table}

\begin{figure*}[ptb]
\begin{center}
\vspace{-3cm}
\includegraphics*[scale=0.7]{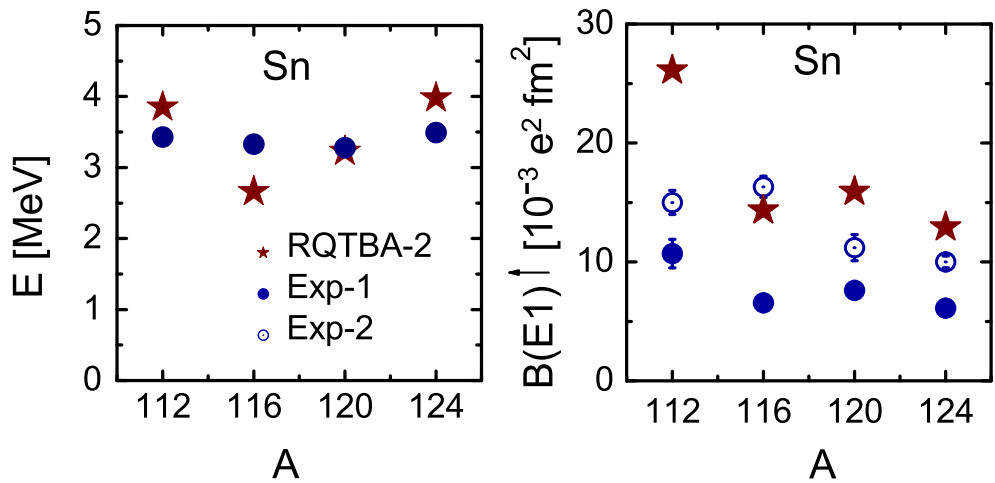}
\caption{Energies and B(E1)$\uparrow$ values of the first $1^-$
states in the chain of tin isotopes $^{112,116,120,124}$Sn obtained
within the relativistic two-phonon model (stars), compared to data
from Refs.~\cite{Pys.06,Bry.99,Gov.98} (filled circles), and
\cite{OEN.07,Oezel.phd} (open circles).} \label{be1}
\end{center}
\end{figure*}

The energies, the corresponding $B(E1)\uparrow$ values and the
$R_{\omega}$ values defined as $R_{\omega} =
\omega(1^-_1)/\bigl(\omega(2^+_1)+\omega(3^-_1)\bigr)$ for the
lowest 1$^-$ states in the tin isotopes are listed in
Table~\ref{tab1}. The experimental energies, $B(E1)\uparrow$ and
$R_{\omega}$ values are taken from
Refs.~\cite{Pys.06,Bry.99,OEN.07,Gov.98,Oezel.phd} where they were
extracted from photon scattering data. The results obtained within
QPM are included for comparison. Notice, that the measurements with
larger end point energies for the electron bremsstrahlung result in
larger $B(E1)\uparrow$ values \cite{Gov.98,OEN.07,Oezel.phd}. Fig.
\ref{be1} shows the results obtained for the energies and reduced
transition probabilities B(E1)$\uparrow$ in the chain of tin
isotopes $^{112,116,120,124}$Sn within the relativistic two-phonon
model. The obtained results are compared to the same sets of data as
in the Table~\ref{tab1}. One can see that the RQTBA-2 results for
the $1^-_1$ states in $^{112,116,120,124}$Sn are in a better
agreement to the data obtained with larger end point energies while
the QPM results rather support another data set. For $^{112,124}$Sn,
however, the QPM transition probabilities are too small.

In RQTBA-2, the position of the first 1$^-$ state is basically
defined by the sum of the energies of the lowest 2$^+$ and 3$^-$
phonons. From the Eq. (\ref{phiresc}) one can see that the amplitude
$\bar{\Phi}(\omega)$ consists of the pole terms with the poles at
the energies which are sums of the two phonon energies. Thus, the
energy of the first 1$^-$ state is approximately equal to the sum of
the energies of the lowest 2$^+$ and 3$^-$ phonons with some
relatively small negative correction introduced by the static
residual interaction $\tilde V$. Similar conclusions follow from the
analysis of the experimentally observed energies of the lowest
$2^+$, $3^-$ and $1^-$ states, see a discussion in
Ref.~\cite{Bry.99}. The above mentioned correction is quantified by
the value of $R_{\omega}$, whose deviation from unity characterizes
the two-phonon anharmonicity, see Table~\ref{tab1}. In particular,
in $^{120}$Sn the energies of the $2^+_1$ and $3^-_1$ phonons
calculated within the RQRPA are obtained at 1.48 MeV and 1.90 MeV,
respectively, and the 1$^-_1$ state appears at 3.23 MeV in the
two-phonon approach. Thus, the quality of description of the first
$1^-$ two-phonon state is mainly determined by the quality of
description provided by the RQRPA for the $2^+_1$ and $3^-_1$
phonons, namely, their energies and coupling vertices. In the cases
of vibrational nuclei, these quantities are reasonably well
described by the RQRPA, however, the description could be further
improved by inclusion of correction beyond RQRPA. A relatively small
anharmonicity allows an identification of the first experimentally
observed $1^-$ state as a member of the $2^+_1\otimes 3^-_1$
quintuplet. Theoretically, we have shown that this state appears
solely due to the inclusion of the two-phonon correlations and does
not appear in the spectra calculated within the conventional RQTBA,
although 2q$\otimes$phonon "prototypes" of this two-phonon state are
present in the RQTBA model space at higher energies. One can see
that for the lowest $1^-$ state in the considered chain of even-even
tin isotopes the obtained agreement of the RQTBA-2 results with the
available data is very good in spite of the fact that this tiny
structure at about 3 MeV originates by the splitting-out from the
very strong RQRPA pygmy mode located at the neutron threshold, due
to the two-phonon correlations included consistently without any
adjustment procedures.
\begin{figure*}[ptb]
\begin{center}
\vspace{-2.5cm}
\includegraphics*[scale=0.9]{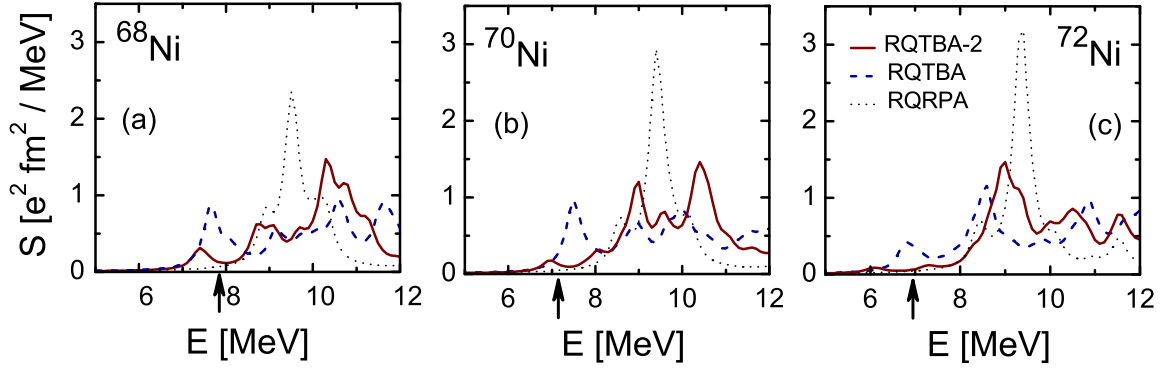}
\vspace{-6.5cm}
\end{center}
\caption{Low-lying dipole spectrum of $^{68,70,72}$Ni calculated
within the RQRPA (black dotted curves), RQTBA (blue dashed curves),
and RQTBA-2 (red solid curves) with a smearing parameter of 200 keV.
The arrows show the neutron thresholds.}
\label{ni}%
\end{figure*}
\begin{figure*}[ptb]
\begin{center}
\vspace{-2.5cm} 
\includegraphics*[scale=0.9]{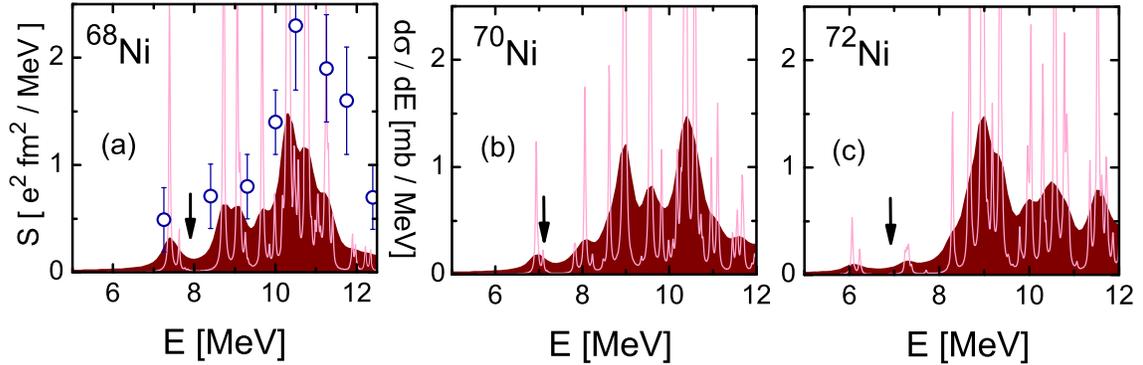}
\vspace{-6.5cm}
\end{center}
\caption{Low-lying dipole spectra of $^{68,70,72}$Ni calculated
within RQTBA-2 with a smearing parameter of 20 keV (thin curves,
light pink) and 200 keV (filled area). Panel (a) contains also the
data from Ref.~\cite{WBC.09} (open circles, units on the right). The
arrows show the neutron thresholds.}
\label{ni200}%
\end{figure*}
The low-lying electromagnetic dipole strengths in neutron rich
$^{68,70,72}$Ni isotopes are displayed in
Figs.~\ref{ni},\ref{ni200}. Fig. \ref{ni} shows the RQTBA-2 strength
functions compared to the RQTBA and the original RQRPA, and Fig.
\ref{ni200} shows the RQTBA-2 strength distributions calculated with
different smearing parameters so that the fine structure of the
strength can be analyzed. In contrast to the case of tin isotopes,
in the considered Ni isotopes almost the whole pygmy dipole
resonance is located above the neutron threshold. For all three
isotopes we found a redistribution and weaker fragmentation of the
low-lying strength in RQTBA-2 compared to the RQTBA calculations of
Ref.~\cite{LRTL.09}. The calculated strength distribution in
$^{68}$Ni has its maximum at 10.30 MeV and the total strength below
12 MeV is 2.73 e$^2$ fm$^2$, while the corresponding fraction of the
energy weighted sum rule (EWSR) is 7.8\% of the total integrated
photoabsorption cross section and 11\% of the Thomas-Reiche-Kuhn sum
rule. These characteristics are in agreement with those extracted
from the recent data reported in Ref.~\cite{WBC.09} which are also
included in Fig. \ref{ni200}. The data are given for the
differential cross section, so that the shape of the measured
distribution can be compared to the theoretical strength. The
RQTBA-2 dipole strength distributions for the $^{70,72}$Ni isotopes
can be suggested as predictions for possible future measurements in
these nuclei. One can see that with the increase of the neutron
number the total strength increases and the centroid of the pygmy
dipole resonance shifts toward lower energy, however, this evolution
is very smooth within the considered isotopic chain. The reason is
the occupation of the 1g$_{9/2}$ intruder orbit in the neutron
subsystem, as it has been discussed in Ref. \cite{LRTL.09}.

By construction, the relativistic two-phonon model represents a step
forward as compared to the conventional RQTBA which includes up to
2q$\otimes$phonon configurations. The inclusion of the additional
two-phonon correlations in the two-quasiparticle propagators
provides a further improvement for the description of the fine structure of
the low-lying strength. Additional correlations between the
quasiparticles redistribute the strength functions because the poles
of the essentially different two-phonon character appear in the
two-quasiparticle propagator. Notice, that the physical content of
the two-phonon RQTBA reminds the two-phonon quasiparticle-phonon
model (QPM) \cite{Sol.92}, however, a one-to-one correspondence
between these models has not been established yet and has to be a
subject of further considerations.

Both RQTBA and RQTBA-2 models are limited by 4q configurations and,
thus, represent, conceptually, the same level of description in
general terms of the configuration complexity. In the RQTBA, each
two-quasiparticle configuration couples to 2q$\otimes$phonon states
and in RQTBA-2 the latter contains the additional coupling between
the 2q to a phonon forming a phonon$\otimes$phonon configuration.
Thus, on the same 4q level of configuration complexity the above
mentioned two types of coupling appear, containing phonon degrees of
freedom. Most probably, the differences between RQTBA and RQTBA-2
results occur because of their limitations in terms of the
configuration space. On a higher level of configuration complexity
involving six and more quasiparticles the differences between the
coupling schemes are expected to be less pronounced. This will be
clarified in the future studies.

\section{Summary}
Summarizing, the two-phonon version of the relativistic time
blocking approximation is presented. Within this model, it has been
shown how the RQRPA modes are fragmented due to the coupling to
two-phonon configurations, thus explaining, in particular, the
physical connection between the pygmy dipole mode and the 1$^-$
member of the $2^+_1\otimes 3^-_1$ quintuplet. A very reasonable
description of the lowest 1$^-$ states in $^{112,116,120,124}$Sn has
been achieved within a fully consistent approach without introducing
any adjustable parameters. The resulting low-lying dipole spectra in
$^{116,120}$Sn are compared with conventional RQTBA calculations. It
is found that the two-phonon correlations redistribute the
fragmented strength as compared to the 2q$\otimes$phonon RQTBA so
that the major fraction of the RQRPA pygmy mode is pushed above the
neutron threshold and therefore mixed with the giant dipole
resonance tail. The calculated low-energy fraction of the
electromagnetic dipole strength agrees also very well with the
available data for the tin isotopes and for the recently
investigated neutron-rich nucleus $^{68}$Ni. In general, the
relativistic two-phonon model presented here provides a new quality
of understanding of mode coupling mechanisms in nuclei and opens the
way for inclusion of higher-order correlations in the nuclear
response function. As the method is based on Green's function
techniques, it can be widely applied also in other areas of quantum
many-body physics.

\bigskip\leftline{\bf ACKNOWLEDGEMENTS}
Fruitful discussions with V. Zelevinsky are gratefully acknowledged.
This work is supported by the National Superconducting Cyclotron
Laboratory at Michigan State University, by the DFG cluster of
excellence \textquotedblleft Origin and Structure of the
Universe\textquotedblright\ (www.universe-cluster.de), and by St.
Petersburg State University under Grant No. 11.38.648.2013.
%
%

\end{document}